\documentclass{article}

\usepackage{arxiv}			
\usepackage[affil-sl]{authblk}
\usepackage[utf8]{inputenc} 
\usepackage[T1]{fontenc}    
\usepackage{hyperref}       
\usepackage{url}            
\usepackage{booktabs}       
\usepackage{amsfonts}       
\usepackage{amsmath}
\usepackage{nicefrac}       
\usepackage{microtype}      
\usepackage[labelfont=bf]{caption}	
\usepackage{graphicx}
\usepackage[numbers]{natbib}
\usepackage{doi}
\usepackage{SIunitx}
\usepackage[section]{placeins}

\date{March 9, 2026}
\title{Large differential attosecond delays in solid state photoemission}

\author[1]{Andreas Gebauer}
\author[1]{Walter Enns}
\author[2]{Sergej Neb}
\author[1]{Tillmann Schabbehard}
\author[1]{Luis Maschmann}
\author[3,4]{Stefan Muff}
\author[3,4]{J. Hugo Dil}
\author[1]{Ulrich Heinzmann}
\author[5,6,7]{Stephan Fritzsche}
\author[8,9]{Ricardo Diez Mui\~{n}o}
\author[8,9,10]{Pedro M. Echenique}
\author[8,11]{Nikolay M. Kabachnik}
\author[8,9,10]{Eugene E. Krasovskii}
\author[1]{Walter Pfeiffer}

\affil[1]{Faculty of Physics, Bielefeld University, Universit\"{a}tsstr. 25, 33615 Bielefeld, Germany}
\affil[2]{Department of Physics, ETH Z\"{u}rich, 8093 Z\"{u}rich, Switzerland}
\affil[3]{Center for Photon Science, Paul Scherrer Institut, CH-5232 Villigen, Switzerland}
\affil[4]{Institute of Physics, \'{E}cole Polytechnique F\'{e}d\'{e}rale de Lausanne, CH-1015 Lausanne, Switzerland}
\affil[5]{Helmholtz-Institut Jena, Fr\"{o}belstieg 3, 07743 Jena, Germany}
\affil[6]{GSI Helmholtzzentrum für Schwerionenforschung GmbH, Planckstr. 1, 64291 Darmstadt, Germany}
\affil[7]{Theoretisch-Physikalisches Institut, Friedrich-Schiller-Universit\"{a}t Jena, Max-Wien-Platz 1, 07743 Jena, Germany}
\affil[8]{Donostia International Physics Center (DIPC), 20018 Donostia – San Sebasti\'{a}n, Spain}
\affil[9]{IKERBASQUE, Basque Foundation for Science, 48013 Bilbao, Spain}
\affil[10]{Universidad del Pa\'{i}s Vasco – Euskal Herriko Unibertsitatea, 20080 Donostia – San Sebasti\'{a}n, Spain}
\affil[11]{European XFEL GmbH, Holzkoppel 4, 22869 Schenefeld, Germany}

\hypersetup{
	pdftitle={Large differential attosecond delays in solid state photoemission},
	pdfsubject={physics.optics, cond-mat.other},
	pdfauthor={Andreas Gebauer, Walter Enns, Sergej Neb, Tillmann Schabbehard, Luis Maschmann, Stefan Muff, J. Hugo Dil, Ulrich Heinzmann, Stephan Fritzsche, Ricardo Diez Muino, Pedro M. Echenique, Nikolay M. Kabachnik, Eugene E. Krasovskii, Walter Pfeiffer},
	pdfkeywords={},
}

\begin{document}
	\maketitle
	
	\begin{abstract}
		Time-resolved photoelectron spectroscopy provides access to the electronic structure and non-equilibrium electron dynamics in matter. At solid surfaces photoemission dynamics can be investigated on its natural time scale by measuring attosecond time delays of emitted electrons \cite{Cavalieri2007}. Photoelectrons with final state energies of several tens of eV need tens to hundreds of attoseconds to be released into the vacuum \cite{Ossiander2018}. Competing effects determine the emission dynamics and, hence, the full picture of the process is still under debate \cite{Tao2016, Siek2017, Kasmi2017, Riemensberger2019, Potamianos2024}. The rather large energy differences between the final states probed in commonly reported relative photoemission delays obscure their complex fine structure and hinders the interpretation of the measurements. Here we report differential attosecond delays $\tau_{\mathrm{DAD}}$, i.e., relative photoemission delays for energetically close-lying spin-orbit split states. Differential attosecond delays on the order of \qtyrange{30}{100}{\as} for Bi 5d, Te 4d, and Se 3d core level photoemission from Bi$_2$Te$_3$ and Bi$_2$Se$_3$ can neither be attributed to intra-atomic delays \cite{Siek2017}, nor to ballistic transport and subsequent emission. Instead, calculations based on the one-step photoemission theory \cite{Kuzian2020} reveal that photoemission delays vary strongly on the energy scale of the spin-orbit splitting and quantitatively match experimental observations. This strong variation arises from multiple scattering at the surface leading to final states that involve both evanescent and propagating Bloch waves. Their relative amplitudes vary strongly affecting thereby the timing of the photoemission event since evanescent and propagating components exhibit inherently different dynamics.
	\end{abstract}
	
	\section*{Introduction}
	
	The emergence of attosecond XUV pulses \cite{Drescher2001, Hentschel2001, Paul2001} enabled the observation of photoemission delays in atoms \cite{Schultze2010}, molecules \cite{Sansone2010} and solids \cite{Cavalieri2007} directly in the time-domain and has yielded results that challenge our understanding of the photoemission process and its accurate theoretical description. The most common methods of attosecond time-resolved photoemission spectroscopy, i.e., attosecond streaking \cite{Itatani2002, Kienberger2002} and reconstruction of attosecond beating by interference of two-photon transitions (RABBITT) \cite{Paul2001, Muller2002}, determine relative photoemission delays between different emission channels. To a large extent measured relative photoemission delays in solids were interpreted as propagation times due to ballistic motion of wave packets \cite{Cavalieri2007, Ossiander2018, Tao2016, Siek2017, Kasmi2017, Riemensberger2019, Neppl2012, Neppl2015, Locher2015, Heinrich2021}, where the phenomenological parameters such as group velocity, effective mass and final state lifetime were determined from the bulk band structure. For solid state photoemission it was shown, that on this level of description delays obtained by RABBITT and streaking are equivalent \cite{Gebauer2019}. Such time-resolved studies provide substantial insights into photoelectron dynamics. In particular it was revealed that there is an influence of resonances \cite{Tao2016, Riemensberger2019, Borisov2013}, final state band gaps \cite{Potamianos2024}, plasmonic excitations \cite{Lemell2015}, electron screening \cite{Chen2017} and intra-atomic many-body interactions \cite{Siek2017} on the photoelectron dynamics. In comparing delays in photoemission from core levels and the valence band \cite{Cavalieri2007} or from core levels of different elements in a compound \cite{Siek2017}, one must take into account that the emission is affected by substantially different final states, so the measured delay depends on various factors, which generally cannot be disentangled. Here we show that for differential attosecond delays, these ambiguities in their interpretation can be sufficiently reduced and conclude that only the rigorous one-step photoemission theory, which explicitly accounts for the scattering of the emitted electron at the surface, can reproduce the observed delays. The trick relies on measuring the relative photoemission delay for photoemission channels that are as close as possible in energy and, in addition, involve final states reached from initial states with the same spatial symmetry, i.e., orbital angular momentum.
	
	The photoelectron dynamics inside the solid is determined by elastic and inelastic scattering in the bulk as well as at the bulk-vacuum interface. Bulk scattering is accounted for in the band structure, and in the most common conception of photoemission the photoelectron is travelling across the crystal with its group velocity $v_{\mathrm{g}}$ and is either washed out of the measurement channel by an inelastic collision or is emitted into the vacuum half space. The short inelastic mean free path of photoelectrons excited by XUV photon energies \cite{Tanuma2011} restricts the emission depth of observable photoelectrons to a few \AA. In this ballistic propagation picture, the photoelectron escape time $t_{\mathrm{esc}}$ is given by $t_{\mathrm{esc}} = \Lambda ⁄ v_{\mathrm{g}}$, where $\Lambda$ is the photoelectron inelastic mean free path. Since $\Lambda = v_{\mathrm{g}}  \tau_{\mathrm{inel}}$, with $\tau_{\mathrm{inel}}$ being the photoelectron lifetime, $v_{\mathrm{g}}$ cancels and $t_{\mathrm{esc}}$ directly corresponds to $\tau_{\mathrm{inel}}$ \cite{Krasovskii2016}. The excited electron lifetime $\tau_{\mathrm{inel}}$ is mostly determined by the available phase space for electron-electron scattering and thus varies only moderately and smoothly with final state energy for energies far above the Fermi level \cite{Silkin2018}. Accordingly, large relative photoemission delays are only expected for large differences in final state energy. However, this equivalence between $t_{\mathrm{esc}}$ and $\tau_{\mathrm{inel}}$ holds only if the group velocity is a meaningful concept. Both for vanishing $v_{\mathrm{g}}$, for example at band edges \cite{Potamianos2024}, and for photoelectrons being emitted via evanescent final states, i.e., final states in band gaps, this simple relation may break down. This was demonstrated for a one-dimensional Kronig-Penney model, which revealed strong variations of $t_{\mathrm{esc}}(E_{\mathrm{kin}})$ in the vicinity of band gaps and close to minima in the photoemission matrix element, i.e., at Cooper minima \cite{Krasovskii2011, Krasovskii2016}. Here we provide experimental evidence that such strong variations of $t_{\mathrm{esc}}(E_{\mathrm{kin}})$ indeed occur and show that the measured differential attosecond delays $\tau_{\mathrm{DAD}}$ can be attributed to the interplay between propagating and evanescent photoemission channels. This relies, both on the ability to measure differential attosecond delays and a quantitative theoretical model that accounts for all effects that are responsible for strong variations of $t_{\mathrm{esc}}(E_{\mathrm{kin}})$. The latter is achieved by employing the one-step photoemission theory \cite{Adawi1964, Mahan1970, Caroli1973, Feibelman1974, Pendry1976}, which, based on the Eisenbud-Wigner-Smith time delay formalism, has been shown to account for the effects of band edges, evanescent emission channels in bulk band gaps, and Cooper minima \cite{Kuzian2020}. 
	
	One-step photoemission theory fully accounts for scattering at the bulk-vacuum interface \cite{Adawi1964, Mahan1970, Caroli1973, Feibelman1974, Pendry1976}. It describes the "click" an emitted free photoelectron causes at a detector as a transition from the ground state to a scattering state: The asymptotic photoelectron wave function at the detector is given by a plane wave whose complex amplitude is determined by the transition matrix element between the ground state and a time-reversed low-energy electron diffraction (TR LEED) state \cite{Krasovskii1999}. The TR LEED state is governed by the elastic and inelastic photoelectron scattering inside the solid and at the surface. This results in a structured final state continuum being distinct from bulk final states in an infinite periodic crystal. If, for example, the final state energy lies in a bulk continuum band gap, the photoelectron is, nevertheless, emitted due to evanescent waves. Thus, the dynamics inside the material and all interfering quantum pathways that as a result of a photoexcitation could lead to the emission of an electron are accounted for. 
	
	\section*{Concept of differential attosecond delays}
	
\begin{figure}
		\centering
		\includegraphics[width = \textwidth]{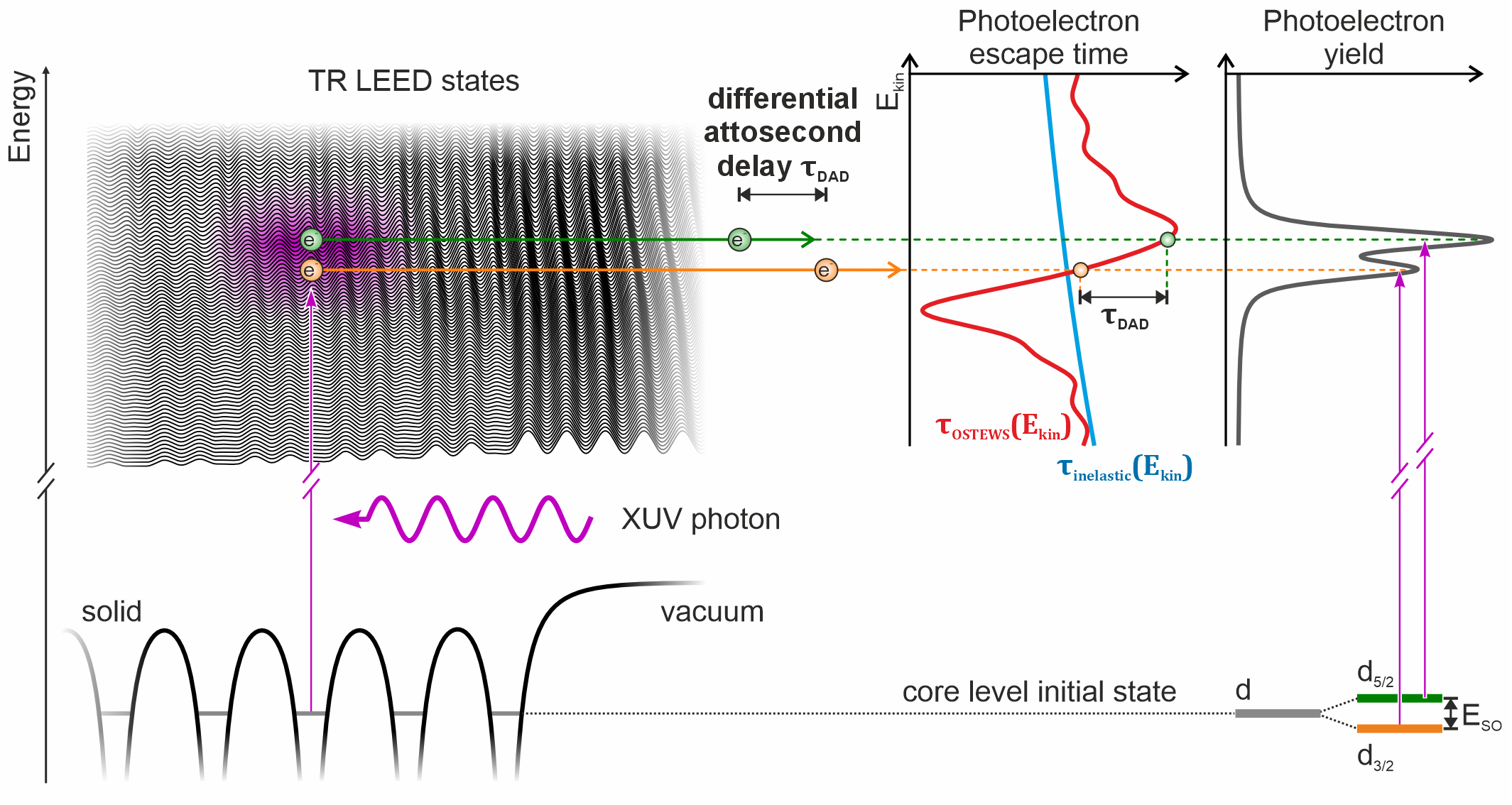}
		\caption{{\bfseries Determination of differential attosecond delays $\boldsymbol{\tau_{\mathrm{DAD}}}$ using spin-orbit split initial states.} An XUV pulse excites photoelectrons from an atom-like initial state subject to spin-orbit splitting with the energy difference $E_{\mathrm{SO}}$ here indicated for d-type initial core levels. The two photoemission final states (indicated by green and orange arrows) differentially probe $t_{\mathrm{esc}}(E_{\mathrm{kin}})$ and yield $\tau_{\mathrm{DAD}}$. For one-step photoemission theory the TR LEED states are indicated as thin black lines in the continuum. They are derived for a one-dimensional finite Kronig-Penney model in the vicinity of a bulk band gap. The Kronig-Penney model parameters are chosen to mimic the 1D single-electron potential shown in the lower part. The $\tau_{\mathrm{OSTEWS}}(E_{\mathrm{kin}})$ derived within one-step photoemission theory using the Eisenbud-Wigner-Smith delay formalism \cite{Kuzian2020} is shown as red solid line in the left inset together with the inelastic electron lifetime $\tau_{\mathrm{inel}}(E_{\mathrm{kin}})$ (blue dotted line), which varies only weakly with energy. The emission from spin-orbit split initial states leads to two spectral contributions in the photoelectron spectrum (right inset panel). Their relative spectral weight reflects the initial state degeneracy.}
		\label{fig:Fig1}
	\end{figure}

	The escape time of photoelectrons can be rigorously calculated by a combination of the one-step theory of photoemission \cite{Adawi1964, Mahan1970, Caroli1973, Feibelman1974, Pendry1976, Krasovskii1999} and the Eisenbud-Wigner-Smith (EWS) time delay formalism (OSTEWS) \cite{Kuzian2020}. Recently, it was shown that the photoelectron escape time determined in streaking spectroscopy agrees with the OSTEWS delay \cite{Kuzian2020}
	\begin{equation}
		\tau_{\mathrm{OSTEWS}} = \hbar \frac{d}{dE} \mathrm{arg}\left( \langle \Phi_{\mathrm{LEED}}^* | \hat{H}_{\mathrm{XUV}} | \psi_0 \rangle \right),
		\label{eq:Eq1}
	\end{equation}
	where $\psi_0$ is the initial state wave function, $\Phi_{\mathrm{LEED}}^*$ is the TR LEED state, and $\hat{H}_{\mathrm{XUV}}$ is the dipole operator. Figure \ref{fig:Fig1} schematically depicts the XUV-driven photoemission for spin-orbit split initial states. For illustration, in the photoemission continuum TR LEED states for final state energies close to a band gap in the continuum are depicted as obtained for a finite Kronig-Penney 1D model assuming inelastic scattering within the solid and no spin-orbit splitting. The corresponding photoelectron escape time calculated based on eq. \ref{eq:Eq1} shows a strong variation with the final state energy (red solid line in left inset in Fig. \ref{fig:Fig1}), whereas the ballistic propagation model predicts an almost constant escape time given by the inelastic lifetime of the photoelectron $\tau_{\mathrm{inel}}(E_{\mathrm{kin}})$ (blue dotted line in left inset in Fig. \ref{fig:Fig1}). For some final state kinetic energies, the photoelectrons are almost simultaneously emitted with the XUV excitation, whereas for other energies the emission is significantly delayed with respect to $\tau_{\mathrm{inel}}(E_{\mathrm{kin}})$. Hence, accounting for scattering with the bulk-vacuum interface within the one-step formalism causes a non-trivial energy dependence of $t_{\mathrm{esc}}(E_{\mathrm{kin}})$. As indicated in Fig. \ref{fig:Fig1} this variability leads to large differential attosecond delays $\tau_{\mathrm{DAD}}$ compared to almost vanishing $\tau_{\mathrm{DAD}}$ given by the slope of $\tau_{\mathrm{inel}}(E_{\mathrm{kin}})$ if the ballistic photoelectron propagation in an absorbing medium is considered.
	
	This rapid energy variation of $t_{\mathrm{esc}}$ can only be probed by measuring the photoemission delays for channels that lie very close in energy, which is achieved by using spin-orbit split initial states. The splitting $E_{\mathrm{SO}}$ of the shallow d$_{5/2}$ and d$_{3/2}$ core levels in the topological insulators Bi$_2$Te$_3$ and Bi$_2$Se$_3$ is on the order of \qtyrange{1}{3}{\eV} \cite{Thompson2001} and thus allow differential probing of $t_{\mathrm{esc}}(E_{\mathrm{kin}})$ on this energy scale. Note, that both photoelectron emission channels originate from the same emitter atom, and since the orbital angular momentum determines the selection rules, basically the same final states are populated in a photoexcitation. For such a small difference in kinetic energy any contribution from semi-classical ballistic propagation is almost completely eliminated. Hence, substantial differential attosecond delays in solids can only arise from a non-trivial energy-dependence of the photoelectron escape time as it is expected in the vicinity of band gaps in the continuum and Cooper minima \cite{Krasovskii2016}. Differential attosecond delays, i.e., relative photoemission delays for spin-orbit split initial states, allow probing the derivative of $t_{\mathrm{esc}}(E_{\mathrm{kin}})$ with simultaneously high temporal and energy resolution. 
	
	\section*{Measurement of differential attosecond delays}
	
	The determination of relative photoemission delays requires to spectrally resolve the spin-orbit splitting in the attosecond time resolved spectroscopy. Attosecond streaking spectroscopy, which relies on the use of an isolated attosecond pulse (IAP) and the interaction of the emitted electrons with the field of a single NIR pulse, would completely wash out the spin-orbit splitting because of the large IAP spectral bandwidth of several eV. Employment of an attosecond pulse train (APT), as it is used in RABBITT spectroscopy, overcomes this limitation. The proper choice of the APT results in a spectrum of distinct harmonics with sufficiently small bandwidth. Still the sidebands induced by the near infrared (NIR) pulse for the spin-orbit split states overlap. Hence, for the determination of differential attosecond delays we rely on a recently demonstrated method for the evaluation of RABBITT spectrograms that yields reliable delays even for highly convoluted spectra, i.e., spectral features for which emission channels corresponding to different harmonics in an attosecond pulse train (APT) and NIR-field induced sideband overlap \cite{Gebauer2025}. 
	
	\begin{figure}
		\centering
		\includegraphics[width = \textwidth]{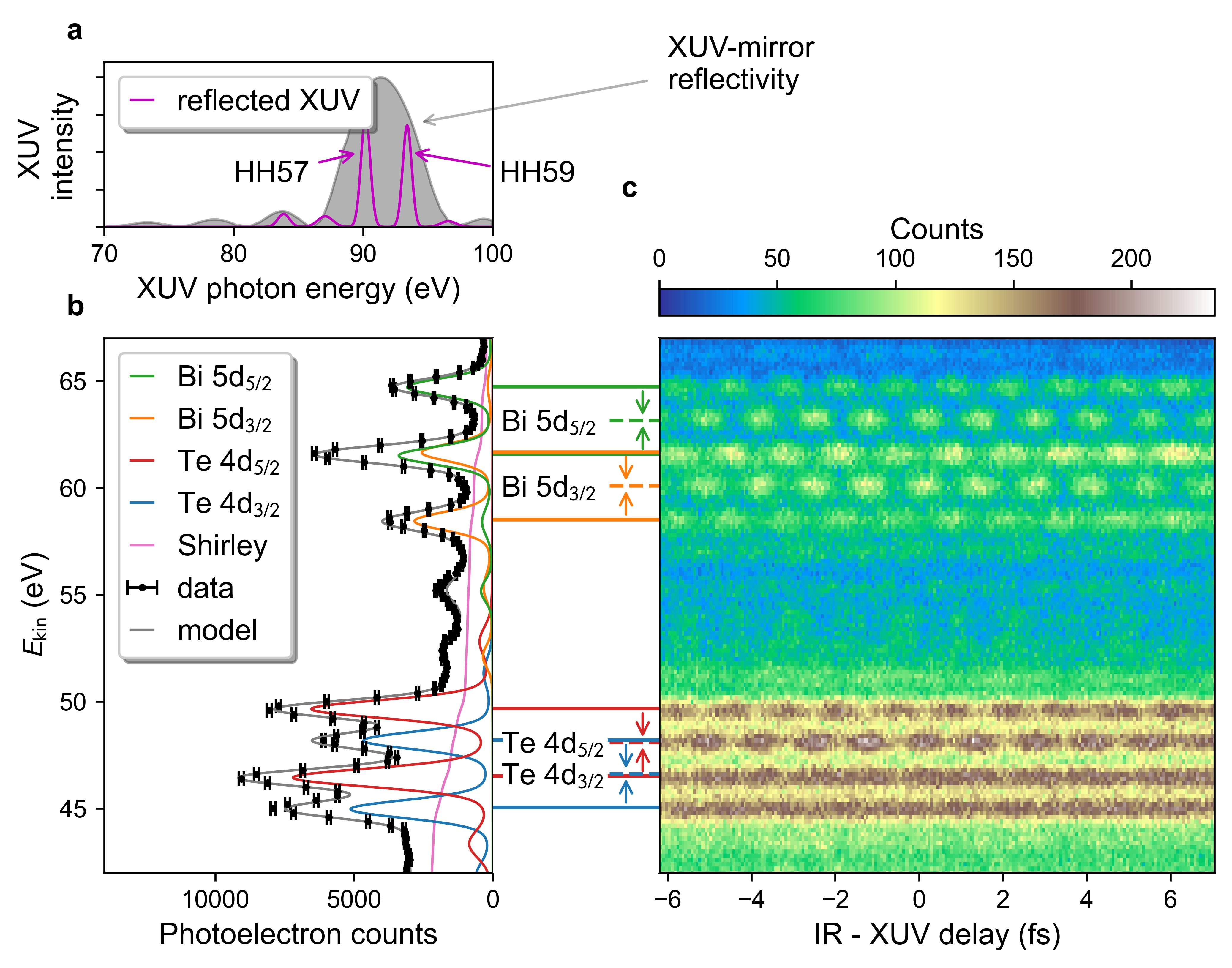}
		\caption{{\bfseries RABBITT measurement for Bi$\boldsymbol{_2}$Te$\boldsymbol{_3}$.} {\bfseries a}, XUV spectrum (magenta) after reflection off a broadband multilayer XUV-mirror with a reflectivity indicated by the gray shaded area. {\bfseries b}, High statistics Bi$_2$Te$_3$ XUV-only photoelectron spectrum (data points) generated by the XUV-spectrum plotted in {\bfseries a}. The error bars correspond to the standard deviation of the underlying Poissonian counting statistics. The fit model function (gray solid line) is the sum of four Voigt profiles for Bi 5d$_{5/2}$ (green), Bi 5d$_{3/2}$ (orange), Te 4d$_{5/2}$ (red) and Te 4d$_{3/2}$ (blue), each convoluted with the XUV spectrum (magenta) shown in {\bfseries a} and the inelastic background calculated as iterative Shirley-Proctor-Sherwood background (pink). {\bfseries c}, RABBITT spectrogram on Bi$_2$Te$_3$. Solid connection lines between {\bfseries b} and {\bfseries c} indicate the principal bands of the respective photoemission channels in the RABBITT spectrogram. The dashed lines indicate the corresponding 58th sidebands with arrows that symbolize the interfering NIR-pulse induced pathways involved.}
		\label{fig:Fig2}
	\end{figure}
	
	The involved XUV attosecond pulse train consists of the 57th ($\hbar \omega_{\mathrm{HH}57} = 90.6 \pm \qty{0.4}{\eV}$) and 59th ($\hbar \omega_{\mathrm{HH}59} = 93.8  \pm \qty{0.4}{\eV}$) harmonics of a NIR fs-laser pulse (Fig. \ref{fig:Fig2}a), where the specified photon energy uncertainty reflects the variation of the harmonics’ peak positions of the complete measurement campaign. With these XUV photon energies photoemission from Bi 5d and Te 4d core levels in Bi$_2$Te$_3$ and Se 3d in Bi$_2$Se$_3$, respectively, is observable. Fig. \ref{fig:Fig2}b shows the photoelectron spectrum of Bi$_2$Te$_3$ excited by the APT (black data points). Details of the analysis of the photoemission spectrum can be found in the methods section. In the RABBITT experiments (Fig. \ref{fig:Fig2}c) the APT and the NIR pulse are brought into spatio-temporal overlap on the sample surface. This leads to laser-assisted photoemission sidebands \cite{Miaja-Avila2006} shifted by the NIR photon energy $\hbar \omega_{\mathrm{NIR}} = \pm 1.59 \pm \qty{0.01}{\eV}$. However, there are two quantum pathways leading to the 58th sideband, i.e. absorption of a photon of the 57th harmonic plus subsequent absorption of an NIR photon at the surface and absorption of a photon of the 59th harmonic plus emission of an NIR photon. Due to quantum path interference with varying relative phase, the photoemission intensity in the 58th sideband oscillates as a function of delay between XUV-APT and NIR pulse \cite{Muller2002,Veniard1996}. In the experimental geometry the NIR field component perpendicular to the surface is strongly suppressed inside the material. Hence, the sideband formation happens only after the photoelectron has been emitted into the vacuum and the resulting RABBITT delays correspond to relative photoelectron escape times $t_{\mathrm{esc}}$. 
	
	The spin-orbit splitting of $E_{\mathrm{SO}}(\mathrm{Bi 5d}) = \qty{3.05}{\eV}$ for the Bi 5d core level leads to a clear separation of the 58th sidebands of Bi 5d$_{3/2}$ at $E_{\mathrm{kin}} = 60.4 \pm \qty{0.4}{\eV}$ and Bi 5d$_{5/2}$ at $E_{\mathrm{kin}} = 63.5 \pm \qty{0.4}{\eV}$, respectively. In the same way the 58th sidebands corresponding to Te 4d$_{3/2}$ at $E_{\mathrm{kin}} = 47.0 \pm \qty{0.4}{\eV}$ and Te 4d$_{\mathrm{5/2}}$ at $E_{\mathrm{kin}} = 48.4 \pm \qty{0.4}{\eV}$ can be identified. Since spin-orbit splitting and NIR photon energy are on the same order of magnitude, Fig. \ref{fig:Fig2}c shows a multitude of overlapping photoelectron peaks in the RABBITT spectrogram. The RABBITT delays are determined by a recently demonstrated method to disentangle overlapping RABBITT spectrograms \cite{Gebauer2025} that combines the complex fit method \cite{Jordan2018} with fixed phase and amplitude relations between sidebands and principal bands in RABBITT spectrograms. 
	
	Evaluation of 213 individual RABBITT measurements on four Bi$_2$Te$_3$ samples leads to a weighted mean value of the Bi 5d differential attosecond delay of $\tau_{\mathrm{DAD}}^{\mathrm{Bi 5d}} = 30 \pm \qty{13}{\as}$, where the error corresponds to the standard deviation of all measurements. Note that the individual delays in all cases are normal distributed and show no trend with the time after cleavage (see Methods section and Supplementary Information Fig. \ref{fig:FigS1} for the statistical distribution of the individual $\tau_{\mathrm{DAD}}$ values). The corresponding Te 4d differential attosecond delay is determined as $\tau_{\mathrm{DAD}}^{\mathrm{Te 4d}} = -39 \pm \qty{18}{\as}$. Literally this means that photoelectrons from the Bi 5d$_{5/2}$ initial state are emitted \qty{30}{\as} later into the vacuum than those from the Bi 5d$_{3/2}$ state, while photoelectrons from Te 4d$_{5/2}$ are emitted \qty{39}{\as} earlier than those from Te 4d$_{3/2}$. Analogous RABBITT measurements on three Bi$_2$Se$_3$ samples yield $\tau_{\mathrm{DAD}}^{\mathrm{Bi 5d}} = 31  \pm \qty{10}{\as}$ and $\tau_{\mathrm{DAD}}^{\mathrm{Se 3d}} = -93 \pm \qty{83}{as}$. The weak Se 3d photoemission yield together with a strong secondary electron background leads to the rather large uncertainty of $\tau_{\mathrm{DAD}}^{\mathrm{Se 3d}}$. For a summary of all determined relative photoemission delays see Supplementary Information Tab. \ref{tab:TabS1}.
	
	\section*{Exclusion of ballistic transport and intraatomic delays}
	
	Photoelectrons from d$_{3/2}$ and d$_{5/2}$ initial states of the same inner shell effectively originate from the same emitter atom and are created with almost the same kinetic energy. Nevertheless, the differential attosecond delays are on the same order of magnitude as relative photoelectron delays in previous studies \cite{Cavalieri2007, Ossiander2018, Tao2016, Siek2017, Kasmi2017, Riemensberger2019, Neppl2012, Neppl2015, Locher2015, Heinrich2021}, which probe relative delays with significantly different final state energies and sometimes different final state continua reached from initial states with different orbital angular momentum. For typical values of the final state lifetime \cite{Silkin2018} a difference in the inelastic lifetime $\tau_{\mathrm{inel}}$ in the order of \qty{30}{\as} would correspond to an unprecedented and unexpected variation of its absolute value by up to about \qty{30}{\percent} for energy differences of a few eV. Moreover, by solving the time-dependent Schrödinger equation for a Jellium model including electron-hole interaction and IR-field penetration into the material \cite{Kazansky2009} or ballistic propagation models we obtain differential attosecond delays for Bi$_2$Te$_3$ and Bi$_2$Se$_3$ that do not exceed three attoseconds for a broad parameter range. Accordingly, all differential attosecond delays that are accessible in the present experiment for Bi$_2$Te$_3$ and Bi$_2$Se$_3$ cannot be explained based on the ballistic picture of photoelectron transport, which predicts negligible relative delays for emission channels with such small final state energy differences.
	
	In addition to the transport of the photoelectron from the emitter atom to the material-vacuum interface intraatomic delays can contribute to $t_{\mathrm{esc}}$. This has been demonstrated for relative photoemission delays in WSe$_2$, which are influenced by the intraatomic centrifugal barrier inside the emitter atom and differ for different initial state angular momenta \cite{Siek2017}. Both the close agreement of the Bi 5d differential attosecond delay of about \qty{30}{\as} in Bi$_2$Te$_3$ and Bi$_2$Se$_3$ and the fact, that for gas phase photoionization emission delays up to several tens of attosecond between spin-orbit-split components in noble gas atoms \cite{Jordan2017, Zhong2020, Ge2021} and molecules \cite{Grafstrom2024} were reported, suggest that also in the present case intraatomic delays could be relevant. In these gas phase experiments the RABBITT NIR probe field acts directly at the location of the emitter atom and the XUV photon energies are small enough to reveal significant spin-orbit delays. In contrast, in solid state photoemission as performed in the present investigation the NIR probe field interacts with the emitted electron only after the emission into the vacuum, i.e., the measured $\tau_{\mathrm{DAD}}$ reflects the intra-atomic processes and the transport induced delays. Since gas phase photoionization experiments for Bi atoms are not feasible we employ relativistic {\itshape ab initio} calculations of the photoionization time-delay for gas phase atoms using the single-configuration Dirac-Fock method \cite{Fritzsche2012}, which accounts for all relativistic effects including spin-orbit interaction. The predicted relative photoionization delays between d$_{5/2}$ and d$_{3/2}$ shells for the various core levels probed are typically smaller than five attoseconds in the photon energy range employed in our experiment. For photon energies similar to those used in the gas phase spin-orbit delay measurements \cite{Jordan2017} our calculations reveal larger relative delays in the order of several tens of attoseconds. This confirms that intraatomic delays could become relevant for small enough photon energies but can be excluded to be responsible for the differential attosecond delays reported here. Hence, the spin-orbit delays determined in atomic gases and the differential attosecond delays reported here are determined by different physical mechanisms. Accordingly, we now exclude both intraatomic delays as well as delays arising from ballistic electron propagation to explain the here reported large values for differential attosecond delays.
	
	\section*{One step theory for $\boldsymbol{\tau_{\mathrm{DAD}}}$}
	
	To overcome the intrinsic limitations of the ballistic propagation model for photoemission we determined $\tau_{\mathrm{DAD}}$ for both materials as function of the XUV photon energy by taking the difference between absolute photoemission delays of d$_{5/2}$ and d$_{3/2}$ photoelectrons that were calculated based on the OSTEWS theory \cite{Kuzian2020} (for details see Methods). Here, we calculated initial and final state wave functions by means of density functional theory (DFT) in local density approximation (LDA). The TR LEED states are calculated by the augmented-plane-wave method \cite{Krasovskii1999, Krasovskii1997}. 
	
	\begin{figure}
		\centering
		\includegraphics[width = \textwidth]{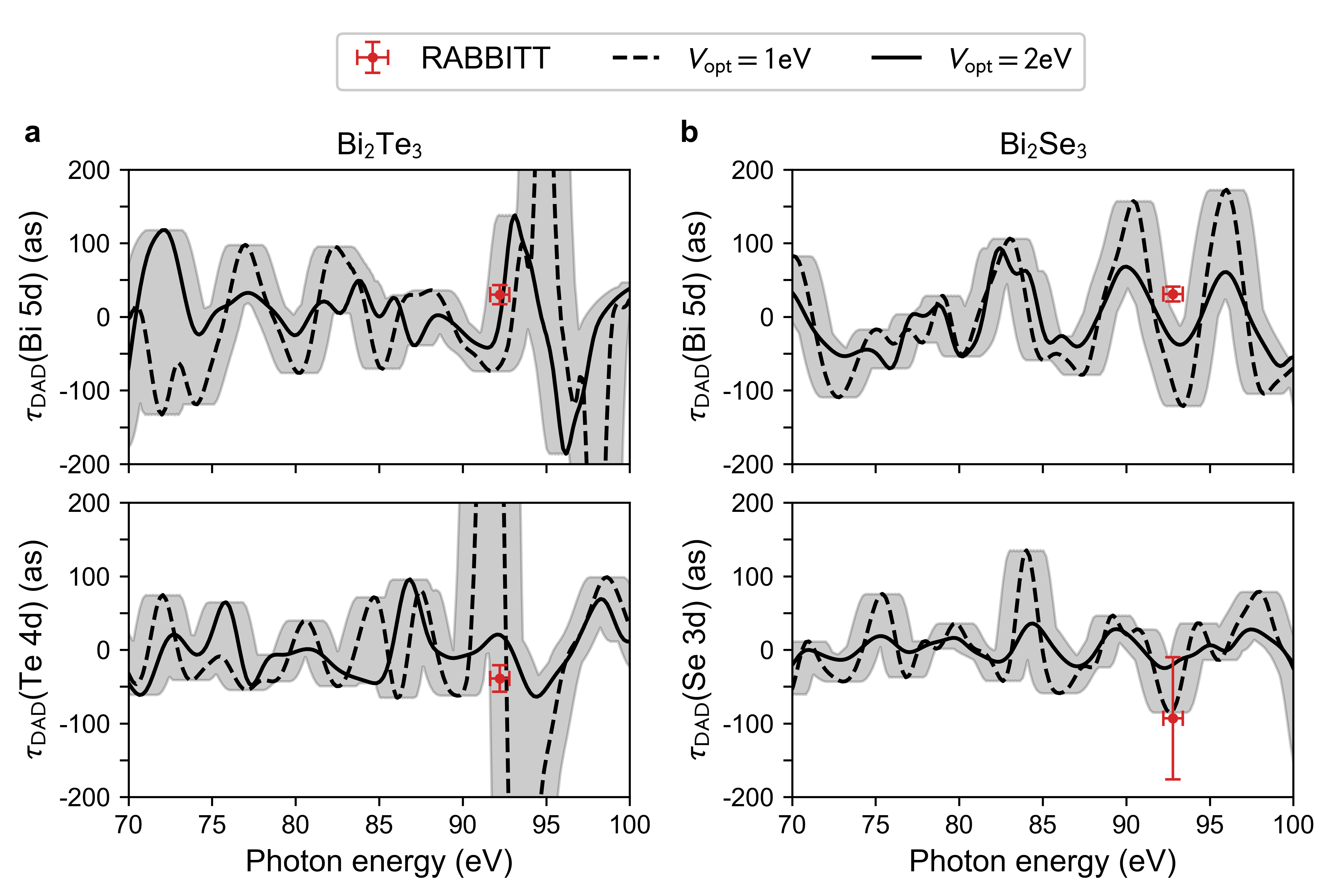}
		\caption{{\bfseries Comparison of theoretical and observed differential attosecond delays.} {\bfseries a} and {\bfseries b}, $\tau_{\mathrm{DAD}}$ as function of the XUV photon energy for Bi$_2$Te$_3$ and Bi$_2$Se$_3$, respectively. The XUV photon energy here corresponds to the even harmonic between the two odd harmonics employed for RABBITT spectroscopy. The black dashed and solid lines correspond to $\tau_{\mathrm{DAD}}$ calculated in the OSTEWS theory assuming $V_{\mathrm{opt}}$ of \qty{1}{\eV} and \qty{2}{\eV}, respectively. The gray shaded areas along each line reflect a $\pm \qty{1}{\eV}$ uncertainty of the absolute energies of the involved Kohn-Sham quasi-particle initial and final state energies derived in LDA based DFT. The data points (in red) are the $\tau_{\mathrm{DAD}}$ observed in the RABBITT experiments. They are determined by the weighted mean values of all experiments and error bars correspond to the weighted standard deviations.}
		\label{fig:Fig3}
	\end{figure}
	
	The determined $\tau_{\mathrm{DAD}}$-curves are shown as function of photon energy for Bi$_2$Te$_3$ and Bi$_2$Se$_3$ in Fig. \ref{fig:Fig3}a and b, respectively. Within the uncertainties of theory and experiment the $\tau_{\mathrm{DAD}}$ values for the four different core-level photoemission channels probed in Bi$_2$Te$_3$ and Bi$_2$Se$_3$ determined via RABBITT spectroscopy (red data points in Fig. \ref{fig:Fig3}) agree quantitatively with the theoretically determined $\tau_{\mathrm{DAD}}$ for the given XUV photon energy. Note that the relative photoemission delays between Bi 5d and Te 4d or Se 3d photoemission channels, respectively, which are measured simultaneously with the $\tau_{\mathrm{DAD}}$, also agree well with the corresponding OSTEWS values (see Supplementary Information Fig. \ref{fig:FigS3}). It indicates that the observed large differential attosecond delays are not a peculiarity of this specific XUV photon energy, but a very general property of solid-state photoemission dynamics, at least for the here studied material systems. Moreover, it suggests that the almost identical differential attosecond delay of Bi 5d in Bi$_2$Te$_3$ and Bi$_2$Se$_3$ is coincidental. Although highly desirable, probing the large variability of $\tau_{\mathrm{DAD}}$ as function of the XUV photon energy is not feasible in the present experimental setup. Hence, the OSTEWS theory accounts for the observed large differential attosecond delays observed in the experiment and, as discussed in the following, provides further insight in the mechanisms that are responsible for such large differential delays.
	
	The OSTEWS $\tau_{\mathrm{DAD}}$ show both large variability as function of the XUV photon energy and $\tau_{\mathrm{DAD}}$ values reaching values well beyond $\pm \qty{200}{\as}$, for example for the Te 4d emission in Bi$_2$Te$_3$. Hence the theory provides evidence that photoemission final states that differ in final state energy in the order of \qtyrange{1}{3}{\eV}, i.e., the spin orbit splitting energy of the initial states, exhibit largely different photoelectron escape times, which are here probed differentially. Additionally, the calculations show that $\tau_{\mathrm{OSTEWS}}$, i.e., the one step theory prediction for $t_{\mathrm{esc}}$, as functions of final state energy are practically identical for d$_{5/2}$ and d$_{3/2}$ photoelectrons (Supplementary Information Fig. \ref{fig:FigS4}). This further confirms that intraatomic delays due to spin-orbit interaction are negligible. Instead, due to their difference in binding energy d$_{5/2}$ and d$_{3/2}$ photoelectrons are excited to different energies in the final state continuum, thereby probing the photoelectron escape time at different final state energies.
	
	The actual values of OSTEWS $\tau_{\mathrm{DAD}}$ are influenced by the choice of the only free parameter in these calculations, i.e., the optical potential $V_{\mathrm{opt}}$, which accounts for the inelastic scattering of the photoelectrons inside the solid. The optical potential is related to $\tau_{\mathrm{inel}}$ of the excited electrons via $V_{\mathrm{opt}} = \hbar ⁄ 2\tau_{\mathrm{inel}}$. The chosen values $1$ eV and $2$ eV correspond to $\tau_{\mathrm{inel}}$ of \qty{329}{\as} and \qty{165}{\as}, respectively. These values lie in the range expected for the given final state energies in the experiment \cite{Krasovskii2016}. The larger $\tau_{\mathrm{inel}}$ applies preferentially for smaller final state energies and vice versa. Note, that an increasing $V_{\mathrm{opt}}$ leads not only to an expected smearing out of spectral structure in the $\tau_{\mathrm{DAD}}$-curves, but also changes features qualitatively, as it is for example obvious at \qty{72}{\eV} photon energy for $\tau_{\mathrm{DAD}}^{\mathrm{Bi 5d}}$ in Bi$_2$Te$_3$ (Fig. \ref{fig:Fig3}a upper panel). A large negative $\tau_{\mathrm{DAD}}$ of about \qty{-100}{\as} changes to an equally large positive $\tau_{\mathrm{DAD}}$ as $V_{\mathrm{opt}}$ increases by \qty{1}{\eV}. This can be understood by the optical potential $V_{\mathrm{opt}}$ and how it differently affects the TR LEED states with predominantly evanescent or propagating character. For vanishing $V_{\mathrm{opt}}$ evanescent TR LEED states decay towards the bulk, whereas propagating TR LEED states are current-carrying and delocalized in the material and exhibit an exponential decay only for finite $V_{\mathrm{opt}}$. Hence, an increase of $V_{\mathrm{opt}}$ increases the relative weight of evanescent emission channels and thus can significantly alter spectral features in the $\tau_{\mathrm{DAD}}$-curves. 
	
	Evanescent waves appear in all types of wave mechanics ranging from acoustics, electromagnetic waves to quantum mechanics. They are characterized by imaginary wave vector components and arise from classically forbidden processes such as for example quantum mechanical tunneling through a potential barrier. The analysis of the group delays of tunneling wave packets or dwell times of particles in potential barriers consistently reveal the advancement of wave packets in tunnel barriers \cite{Hartman1962, Hauge1989, Winful2006}. The availability of ultrashort laser pulses has enabled studying this advancement of evanescent waves directly in the time domain for light propagation in band gaps of photonic crystals \cite{Spielmann1994}, atomic tunnel ionization \cite{Eckle2008, Sainadh2019} and atoms tunneling through an optical barrier \cite{Ramos2020}. Assuming that the same holds for photoemission channels, the large variability seen in OSTEWS $\tau_{\mathrm{DAD}}$ (Fig. \ref{fig:Fig3}) is attributed to the large difference in the emission dynamics for emission channels within band gaps and at band edges. Emission via evanescent TR LEED states in band gaps is advanced, whereas the emission via states at band edges is significantly delayed due to the vanishing group velocity exactly at the band edge. Hence, advancement and increased delay could appear for energetically close lying photoemission final states. Note that the mixing of these effects in case of small band gaps leads to rather complex variations of the escape time \cite{Krasovskii2016}. In particular for the rather complex continuum band structure of the materials studied here an identification of band gaps and band edges is impossible. Still the large variability of OSTEWS $\tau_{\mathrm{DAD}}$ and in particular its variation with the optical potential shows that the two prototypical cases of ballistic and evanescent transport can be qualitatively distinguished and give rise to the large $\tau_{\mathrm{DAD}}$ reported here.
	
	\section*{Conclusions}
	
	In summary, we reported the first experimental observation of differential attosecond delays employing spin-orbit-split core levels as initial states in a solid via RABBITT measurements. They probe the variation of the emission dynamics with unprecedented energy resolution. The reported delays in the order of several tens of attoseconds are of the same magnitude as relative delays between different core level photoelectrons initially localized at different atomic species in a compound material. This observation alone proves that the picture of photoelectrons propagating like nearly-free electrons with well-defined group velocities inside the solid has to be abandoned. To accurately predict photoemission delays it is inevitable to account for the surface-induced breaking of the lattice periodicity. The resulting structured final state continuum consists of evanescent and propagating waves and their largely varying relative weight is responsible for the large differential attosecond delays. Hence, it is the complicated multiple scattering of the outgoing photoelectron in the presence of the surface that determines how long it takes for a photoelectron to be emitted from the solid into the vacuum.
	
	\section*{Methods}
	
	\subsection*{Experimental Setup}
	
	\qty{30}{\fs} near-infrared pulses (\qty{790}{\nm} center wavelength) are generated by a commercial Ti:sapphire-based laser system (Femtopower PRO) at a repetition rate of \qty{3}{\kHz}. The laser pulse is temporally compressed to \qty{18}{\fs} by subsequent spectral broadening via self-phase modulation in a neon-filled hollow core fiber (\qty{1}{\metre} length, \qty{250}{\micro\metre} core diameter, \qty{400}{\milli\bar} Ne gas pressure) and spectral phase control in a chirped mirror compressor. 
	
	XUV pulses are generated via high harmonic generation in a neon gas target inside a vacuum chamber. A differential pumping scheme reduces the base pressure from the HHG target along the beam path by twelve orders of magnitude to \qty{10e-10}{\milli\bar} at the sample position. XUV and NIR pulses propagate collinearly, while the XUV beam diverges much less than the NIR beam. Both beams are spatially separated by a \qty{150}{\nm} thick Zr foil glued on a pellicle iris whose diameter matches the XUV beam diameter. The NIR beam passes outside the Zr foil forming a ring-shaped beam whose intensity on the sample is controlled via a neutral density filter and an iris. Both beams are incident on a co-axial spherical double mirror assembly. The NIR beam is focused by the outer part of the double mirror onto the sample (\qty{125}{\mm} focal length) while the XUV beam is focused by the inner part. The inner part of the double mirror assembly acts as a spectral filter via a Mo-Si multilayer coating with a reflection maximum at \qty{91.2}{\eV} and \qty{5.8}{\eV} bandwidth. It therefore only reflects the 57th and 59th harmonics (Fig. \ref{fig:Fig2}a). This spectral filtering forms attosecond pulse trains (APTs) that consist of exactly two adjacent odd-order harmonics located at $90.6 \pm \qty{0.4}{\eV}$ and $93.8 \pm \qty{0.4}{\eV}$. Moreover, the inner mirror is mounted on a high-precision piezo-driven delay stage that controls the delay between APT and NIR pulse. 
	
	Both pulses are p-polarized and incident on the sample surface in an \ang{85} angle with respect to the surface normal. Photoelectrons are detected in normal emission direction in a field-free time-of-flight spectrometer with an acceptance angle of $\pm\ang{3}$. The spectrometer energy axis was calibrated using Auger lines in the Xe photoelectron spectrum. Residual magnetic fields at the sample position are compensated to \qty{< 1}{\micro T} using three Helmholtz coils. 
	
	Every RABBITT measurement contains 200 delay steps with a delay step size of approximately \qty{70}{\as}. 
	
	\subsection*{Sample preparation}
	
	Bi$_2$Te$_3$ bulk crystals were grown by a collaborator by the Bridgeman method and CVT grown Bi$_2$Se$_3$ bulk crystals were purchased \cite{2DSemiconductors}. The crystals were in-situ cleaved in ultra-high vacuum at a base pressure \qty{< 10e-10}{\milli\bar}. In total 213 individual RABBITT measurements on four different Bi$_2$Te$_3$ samples and 187 RABBITT measurements on three different Bi$_2$Se$_3$ samples were performed. After every RABBITT measurement a high statistics photoelectron spectrum without temporal overlap between XUV and NIR pulses was recorded to observe possible changes in the photoelectron spectrum due to surface contamination. However, no significant changes in photoelectron spectra and delays (Supplementary Information Fig. \ref{fig:FigS1} and Fig. \ref{fig:FigS2}) were observed within the first 144 hours after cleaving, indicating that the cleaved surfaces remained stable for the entire measurement period. All measurements were performed at room temperature. 
	
	\subsection*{Modeling the photoelectron spectrum in absence of the NIR pulse}
	
	To disentangle contributions from different photoemission channels and different harmonic orders within the XUV pulse train, a high-resolution XPS spectrum of a Bi$_2$Te$_3$ sample from the same batch was measured at the Surface/Interface Science beamline of the Swiss Light Source facility with monochromatic synchrotron radiation at a photon energy of \qty{91}{\eV}. In the region of interest four core level photoemission channels in Bi$_2$Te$_3$ (Te 4d$_{3/2}$, Te 4d$_{5/2}$, Bi 5d$_{3/2}$ and Bi 5d$_{5/2}$) could be identified. This XPS spectrum was used to determine binding energies, linewidths and intensity ratios by fitting a Voigt profile to each peak and a two-parameter Shirley-Proctor-Sherwood background \cite{Herrera-Gomez2014}. 
	
	In the analysis of the high statistics XUV-only photoelectron spectrum excited by the APT (Fig. \ref{fig:Fig2}b) all Voigt profile parameters were kept fixed. The sum of the four Voigt profiles is convolved with the XUV spectrum consisting of a series of five Gaussians (Fig. \ref{fig:Fig2}a). The peak positions, widths and intensities of the Gaussians as well as the Shirley-Proctor-Sherwood background parameters were the only free fit parameters in the model function (Gray line in Fig. \ref{fig:Fig2}b), resulting in a good agreement with the measured photoelectron spectrum. 
	
	Since the Bi 5d spin-orbit splitting of \qty{3.05}{\eV} is very close to the energy difference between the two harmonics in the XUV spectrum, i.e. $2\hbar \omega_{\mathrm{NIR}} = \qty{3.18}{\eV}$, a triple peak structure appears in the region of the Bi 5d peaks. However, the Te 4d spin-orbit splitting of \qty{1.47}{\eV} is close to the fundamental photon energy of \qty{1.59}{\eV}, yielding clearly discernible peaks in the high statistics measurement out of temporal overlap with the NIR pulse, but overlapping and almost $\pi$-shifted sideband oscillations in the RABBITT spectrogram (Fig. \ref{fig:Fig2}c). 
	
	Bi$_2$Se$_3$ was treated in a similar way with peak parameters taken from literature \cite{Thuler1982}. 
	
	\subsection*{RABBITT delay extraction}
	
	The RABBITT delays were extracted following an approach based on the complex fit method \cite{Jordan2018} including additional constraints for phase and amplitude relations between sidebands and principal bands that were recently suggested by some of the authors and validated for simulated data \cite{Gebauer2025}. The RABBITT spectrogram raw data was Fourier transformed along the NIR-XUV-delay axis via non-uniform discrete Fourier transform (NUDFT) considering the non-uniformity of the piezo-stage step size. To avoid artifacts from the finite measurement range, the time-domain raw data was multiplied with a Kaiser-Bessel window function with window parameter $\beta = 4$ and zero-padded. The $2f_{\mathrm{NIR}}$ peak was individually determined in every measurement with $f_{\mathrm{NIR}}$ being the NIR pulse center frequency. The frequency domain data was then spectrally filtered at $2f_{\mathrm{NIR}}$ by an eighth-order super-Gaussian with a full width half maximum of \qty{100}{\THz} and integrated along the frequency axis to yield the frequency integrated RABBITT spectrum $Z(E_{\mathrm{kin}})$, i.e.
	\begin{equation}
		Z(E_{\mathrm{kin}}) = \int_{-\infty}^{\infty} I(f, E_{\mathrm{kin}}) e^{-256 \ln (2) \left( \frac{f - 2 f_{\mathrm{NIR}}}{\mathrm{FWHM}} \right)^8} df ,
		\label{eq:Eq2}
	\end{equation}
	with $I(f, E_{\mathrm{kin}})$ being the Fourier transform of the RABBITT photoemission raw data. The influence of window function, zero-padding and spectral filter bandwidth on the RABBITT delays was carefully investigated. It did not shift the mean delay values significantly, but this choice of filters and parameters reduced the standard deviation of data points considerably. For every kinetic energy $Z(E_{\mathrm{kin}})$ is a complex number whose amplitude corresponds to the amplitude of the $2f_{\mathrm{NIR}}$ photoemission intensity oscillation and whose phase describes its phase shift with respect to the origin of the NIR-XUV delay axis.
	
	Making use of both the fixed RABBITT phase and amplitude relations \cite{Gebauer2025}, the frequency filtered and integrated data $Z(E_{\mathrm{kin}})$ is fitted by a model function
	\begin{equation}
		Z_{\mathrm{fit}}(E_{\mathrm{kin}}) = \sum_n \frac{-I_{n, 57}(E_{\mathrm{kin}}) + 2 I_{n, 58}(E_{\mathrm{kin}}) - I_{n, 59}(E_{\mathrm{kin}})}{2} A_{n, 58} e^{i \varphi_{n, 58}} ,
		\label{eq:Eq3}
	\end{equation}
	where $n$ is an index for a specific photoemission channel and $A_{n,58}$ and $\varphi_{n,58} = \frac{2 E_{\mathrm{NIR}}}{\hbar} \tau_{n,58}$ are its RABBITT oscillation amplitude and phase, respectively. $I_{n,q}(E_{\mathrm{kin}})$ is the peak function corresponding to the photoelectron peak associated with photoemission channel $n$ at the $q$-th sideband or principal band. This peak function was determined for every measurement from a high statistics XUV-only photoelectron spectrum as described in the previous section. Therefore, the delay extraction of each photoemission channel relies on only two independent fit parameters for each photoemission channel. The delay difference between two photoemission channels $n$ and $m$ is then given by
	\begin{equation}
		\tau_{n-m} = \hbar \frac{\varphi_{n,58} - \varphi_{m, 58}}{2 E_{\mathrm{NIR}}} .
		\label{eq:Eq4}
	\end{equation}
	The NIR photon energy is determined from the photoelectron peak positions by $\frac{E_{n,59} - E_{n,57}}{2}$. 
	
	To validate the application of the fixed RABBITT phase and amplitude relations, the delay extraction of the Bi 5d differential attosecond delay provides a benchmark because both contributions can be clearly separated. If the fixed relations are dropped, the average $\tau_{\mathrm{DAD}}^{\mathrm{Bi 5d}}$ of all RABBITT measurements on Bi$_2$Te$_3$ yields $24 \pm \qty{12}{\as}$, which agrees within the error bar with the results for fixed phase relations. Instead, the overlapping peaks of Te 4d result in a significantly larger standard deviation of the differential attosecond delay if the phase and amplitude relations are dropped and their application is clearly beneficial. 
	
	\subsection*{Error bars}
	
	We estimated the influence of photoemission counting statistics by taking one RABBITT measurement and generating 100 samples of the same measurement by adding random numbers to it. These random numbers follow Poisson statistics with a variance that is given by the number of photoelectrons detected in an energy channel in the original measurement. Evaluation of these 100 artificial RABBITT data sets revealed that the mean delay values are still equal to the original delay value. The variance of data points across these 100 samples is smaller than the variances of the fit parameters. From these findings we concluded that the photoelectron spectrometer integration time is sufficient to exclude counting statistics as the limiting factor. Moreover, this allows to use the fit parameter variances as estimates for the error bars of the individual measurements. The delay difference error of an individual RABBITT measurement is therefore estimated by Gaussian error propagation from Eq. \ref{eq:Eq4} to
	\begin{equation}
		\Delta \tau_{n-m} = \sqrt{ \left( \frac{\hbar}{2E_{\mathrm{NIR}}} \Delta \varphi_{n,58} \right)^2 + \left( \frac{\hbar}{2E_{\mathrm{NIR}}} \Delta \varphi_{m,58} \right)^2 + \left( \frac{\hbar\left( \varphi_{n,58}  - \varphi_{m,58} \right)}{2E_{\mathrm{NIR}}^2} \Delta E_{\mathrm{NIR}} \right)^2} ,
		\label{eq:Eq5}
	\end{equation}
	with $\Delta \varphi_{n,58}$ being the standard deviation of the $\varphi_{n,58}$ fit parameter and $\Delta E_{\mathrm{NIR}} = \frac{1}{2} \sqrt{\left( \Delta E_{n,57} \right)^2 + \left( \Delta E_{n,59} \right)^2}$, where $\Delta E_{n,q}$ is the fit parameter standard deviation from the XUV-only photoelectron spectrum. 
	All delay differences given in the main text are weighted mean values of 213 RABBITT measurements in case of Bi$_2$Te$_3$ and 187 RABBITT measurements in case of Bi$_2$Se$_3$. They are calculated via
	\begin{equation}
		\langle \tau \rangle = \left. \sum_i^N \frac{\tau_i}{\Delta \tau_i^2} \middle/ \sum_i^N \frac{1}{\Delta \tau_i^2} \right. ,
		\label{eq:Eq6}
	\end{equation}
	where $\tau_i \pm \Delta \tau_i$ is the RABBITT delay of the $i$-th RABBITT measurement and $N$ is the number of measurements. The errors of the weighted mean delays in the main text are given by the weighted standard deviation of the ensemble of all measurements \cite{Bevington}
	\begin{equation}
		\sigma = \sqrt{\frac{N}{N-1} \left( \sum_i^N \frac{\tau_i^2}{\Delta \tau_i^2} \middle/ \sum_i^N \frac{1}{\Delta \tau_i^2} - \langle \tau \rangle^2 \right)} .
		\label{eq:Eq7}
	\end{equation}
	
	\subsection*{Calculation of OSTEWS delays}
	
	The Wigner time delay of the photoelectrons emitted from the semi-core d$_{3/2}$ and d$_{5/2}$ states was calculated within the framework of the one-step theory \cite{Kuzian2020}. The electronic structure of the occupied states was obtained within the density functional theory with the relativistic effects included in second-variational two-component Koelling-Harmon approximation \cite{Koelling1977}. All calculations were performed with the extended augmented plane waves method \cite{Krasovskii1997}. In view of the atomic-like nature of the photoemission initial states they were calculated as eigenstates of a one-formula-unit-thick isolated slab (quintuple layer). The time-reversed LEED states were calculated in the scalar-relativistic approximation for a semi-infinite crystal terminated by a potential step \cite{Krasovskii1999}. The excitation amplitudes were calculated in the dipole approximation as momentum matrix elements for the polarization vector perpendicular to the crystal surface. The energy derivative of the phase of the matrix element yields the photoelectron’s arrival time at a reference plane assuming that the equation of motion of the outgoing wave packet in vacuum can be extrapolated to that plane. In the present calculation the plane was located half the van der Waals gap above the outermost atomic layer. For comparison with experimental data (Fig. \ref{fig:Fig3} and Fig. \ref{fig:FigS3}) the convolution of the OSTEWS delay and a Gaussian with full width half maximum (FWHM) equal to the FWHM of the respective RABBITT sideband peaks is calculated. 
	
	\bibliographystyle{srtnat}

	{\bfseries Acknowledgements:} We gratefully acknowledge Helmuth Berger from EPFL for providing us with Bi$_2$Te$_3$ samples. This work was supported by the German Research Foundation (DFG) within the Priority Program SPP 1840 “Quantum Dynamics in Tailored Intense Fields” (A.G. and W.P., PF 317/10-2, 652733), by the Federal Ministry of Education and Research (A.G., BMBF Grant No. 05K22PBA) and by the Spanish Ministry of Science and Innovation (E.E.K. Projects No. PID2022-139230NB-I00 and No. PID2022-138750NB-C22). N.M.K. acknowledges the hospitality and the financial support from European XFEL and Donostia International Physics Center. 
	
	{\bfseries Author contributions:} A.G. and W.E. performed the experiment. A.G. analyzed the experimental data, calculated Jellium-model photoemission delays and prepared the first manuscript draft. S.N., T.S. and L.M. contributed substantially to the experiment. S.M. and J.H.D. provided XPS spectra of Bi$_2$Te$_3$. S.F. and N.M.K. calculated atomic photoionization delays. E.E.K. calculated the solid-state OSTEWS delays. W.P. coordinated the project. A.G., S.N., T.S., L.M., R.D.M., P.M.E., N.M.K., E.E.K. and W.P. developed the physical interpretation of differential attosecond delays. All authors contributed to the preparation of the manuscript.
	
	{\bfseries Competing Interests:} The authors declare no competing interests.
	
	\appendix
	\newpage
	\section*{Supplementary Information}
	
	The RABBITT experiments not only give access to the differential attosecond delays, but also to all relative delays between each pair of the four photoemission channels (Tab. \ref{tab:TabS1}). These four additional relative delays are plotted as red error crosses in Fig. \ref{fig:FigS3}a and b for Bi$_2$Te$_3$ and Bi$_2$Se$_3$, respectively. Figure \ref{fig:FigS3} also shows the corresponding relative delays obtained from OSTEWS-theory. 
	
	\begin{table}[h!]
		\centering
		\caption{{\bfseries Relative RABBITT delays of all photoemission channels with respect to each other.} All values correspond to the mean values of 213 RABBITT measurements on four Bi$_2$Te$_3$ samples and 187 RABBITT measurements on three Bi$_2$Se$_3$ samples, respectively. The errors correspond to the weighted standard deviation described in the methods section. }
		\label{tab:TabS1}
		{\renewcommand{\arraystretch}{1.5} 
		\begin{tabular}{c|c}
			Bi$_2$Te$_3$ & Bi$_2$Se$_3$ \\ \hline
			\begin{tabular}{c|r}
				$\tau(\mathrm{Bi 5d}_{5/2}) - \tau(\mathrm{Bi 5d}_{3/2})$ & $30 \pm \qty{13}{\as}$ \\ \hline
				$\tau(\mathrm{Te 4d}_{5/2}) - \tau(\mathrm{Te 4d}_{3/2})$ & $-39 \pm \qty{18}{\as}$ \\ \hline
				$\tau(\mathrm{Bi 5d}_{5/2}) - \tau(\mathrm{Te 4d}_{5/2})$ & $18 \pm \qty{15}{\as}$ \\ \hline
				$\tau(\mathrm{Bi 5d}_{3/2}) - \tau(\mathrm{Te 4d}_{3/2})$ & $-20 \pm \qty{22}{\as}$ \\ \hline
				$\tau(\mathrm{Bi 5d}_{3/2}) - \tau(\mathrm{Te 4d}_{5/2})$ & $-13 \pm \qty{16}{\as}$ \\ \hline
				$\tau(\mathrm{Bi 5d}_{3/2}) - \tau(\mathrm{Te 4d}_{3/2})$ & $-53 \pm \qty{23}{\as}$ \\ 
			\end{tabular}
			&
			\begin{tabular}{c|r}
				$\tau(\mathrm{Bi 5d}_{5/2}) - \tau(\mathrm{Bi 5d}_{3/2})$ & $31 \pm \qty{10}{\as}$ \\ \hline
				$\tau(\mathrm{Se 3d}_{5/2}) - \tau(\mathrm{Se 3d}_{3/2})$ & $-93 \pm \qty{83}{\as}$ \\ \hline
				$\tau(\mathrm{Bi 5d}_{5/2}) - \tau(\mathrm{Se 3d}_{5/2})$ & $25 \pm \qty{52}{\as}$ \\ \hline
				$\tau(\mathrm{Bi 5d}_{5/2}) - \tau(\mathrm{Se 3d}_{3/2})$ & $-65 \pm \qty{61}{\as}$ \\ \hline
				$\tau(\mathrm{Bi 5d}_{3/2}) - \tau(\mathrm{Se 3d}_{5/2})$ & $-5 \pm \qty{52}{\as}$ \\ \hline
				$\tau(\mathrm{Bi 5d}_{3/2}) - \tau(\mathrm{Se 3d}_{3/2})$ & $-96 \pm \qty{61}{\as}$ \\ 
			\end{tabular}
		\end{tabular}
		}
	\end{table}
	
	\begin{figure}
		\centering
		\includegraphics[width = 0.9\textwidth]{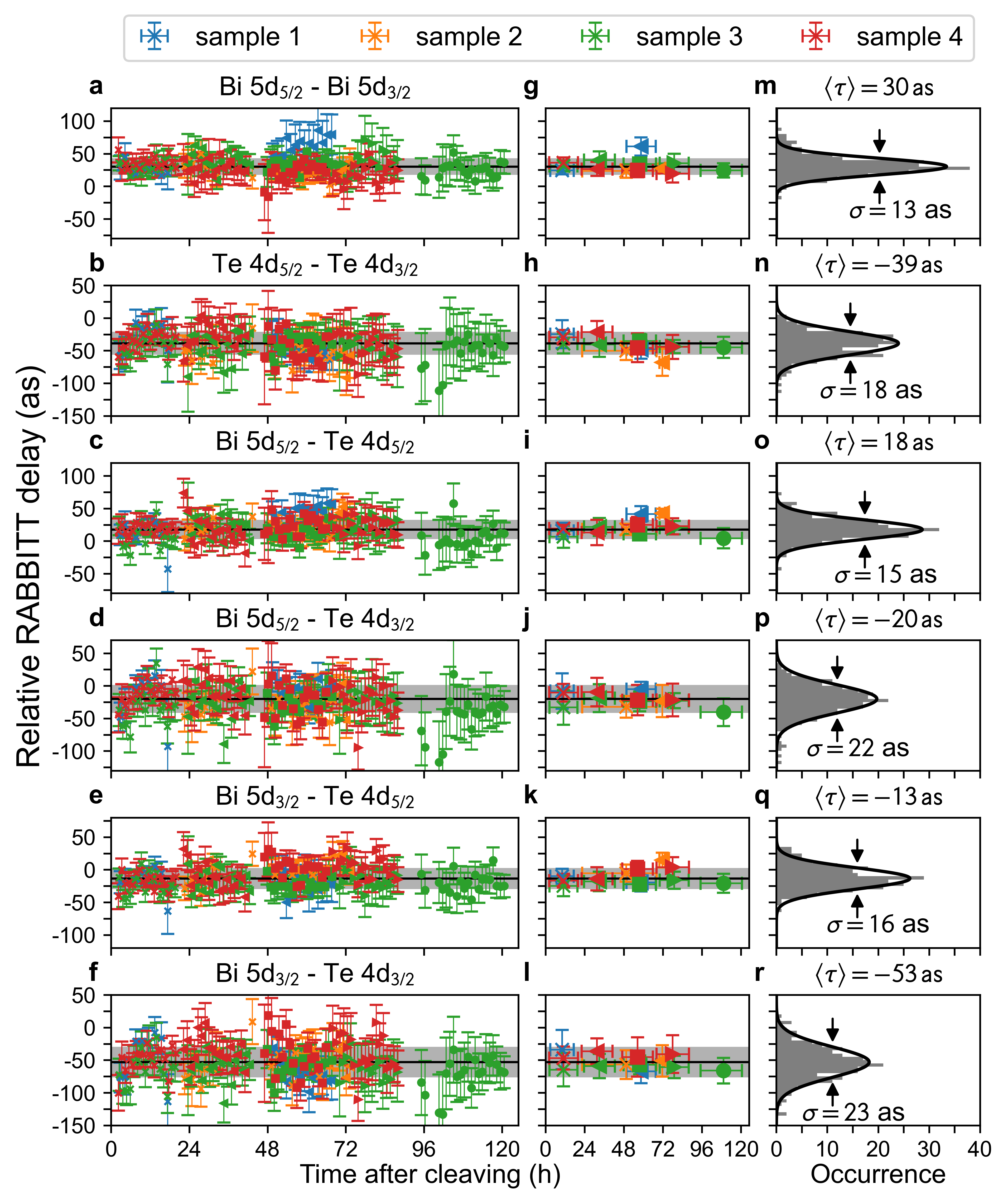}
		\caption{{\bfseries Extracted delays of 213 individual RABBITT measurements on Bi$\boldsymbol{_2}$Te$\boldsymbol{_3}$.} Each row shows a pair of RABBITT delay differences $\tau_n - \tau_m$ between two photoemission channels $n$ and $m$, with $n,m \in \left\{ \mathrm{Bi 5d}_{5/2}, \mathrm{Bi 5d}_{3/2}, \mathrm{Te 4d}_{5/2}, \mathrm{Te 4d}_{3/2} \right\}$. If the difference is positive, photoemission channel $n$ is emitted delayed with respect to channel $m$. {\bfseries a-f}, All extracted delay values as a function of time after cleaving a fresh sample. The color indicates the sample, and the symbol indicates a position on the sample surface where multiple measurements were performed. Horizontal error bars correspond to the duration of the measurement and vertical error bars correspond to Eq. \ref{eq:Eq5}. The black line indicates the weighted mean value and the gray shaded area the weighted standard deviation. {\bfseries g-l}, Weighted mean of the delays of equivalent measurements, i.e. measurements performed in the same position on the sample. {\bfseries m-r}, Occurrence of measured delay values with histogram bins of \qty{5}{\as} width. The black curve indicates a normal distribution corresponding to the weighted mean value $\langle \tau \rangle$ (Eq. \ref{eq:Eq6} in Methods) and weighted standard deviation $\sigma$ (Eq. \ref{eq:Eq7} in Methods).}
		\label{fig:FigS1}
	\end{figure}
	
	\begin{figure}
		\centering
		\includegraphics[width = 0.9\textwidth]{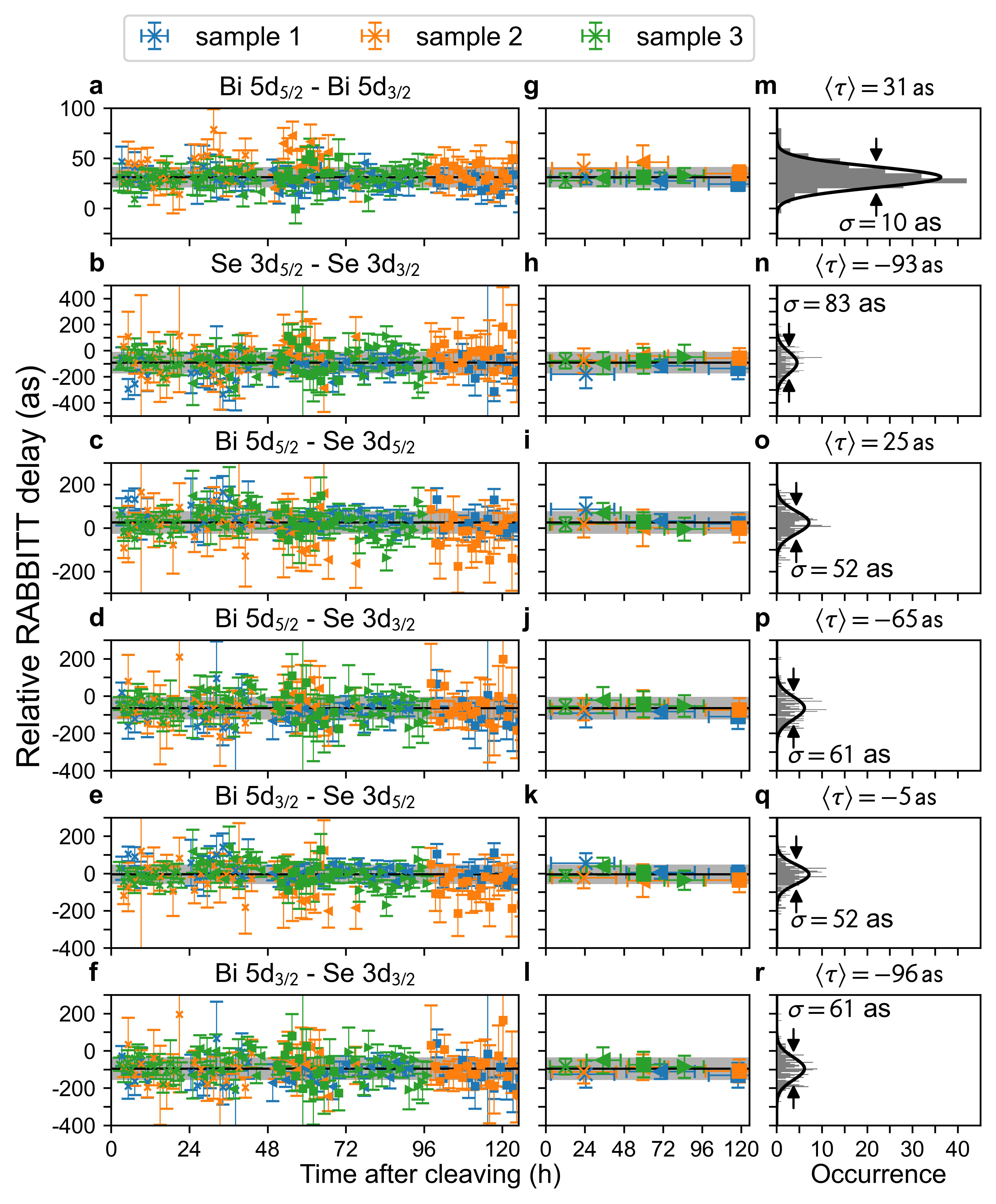}
		\caption{{\bfseries Extracted delays of 187 individual RABBITT measurements on Bi$\boldsymbol{_2}$Se$\boldsymbol{_3}$}. Each row shows a pair of RABBITT delay differences $\tau_n - \tau_m$ between two photoemission channels $n$ and $m$, with $n,m \in \left\{ \mathrm{Bi 5d}_{5/2}, \mathrm{Bi 5d}_{3/2}, \mathrm{Se 3d}_{5/2}, \mathrm{Se 3d}_{3/2} \right\}$. If the difference is positive, photoemission channel $n$ is emitted delayed with respect to channel $m$. {\bfseries a-f}, All extracted delay values as a function of time after cleaving a fresh sample. The color indicates the sample, and the symbol indicates a position on the sample surface where multiple measurements were performed. Horizontal error bars correspond to the duration of the measurement and vertical error bars correspond to Eq. \ref{eq:Eq5}. The black line indicates the weighted mean value and the gray shaded area the weighted standard deviation. {\bfseries g-l}, Weighted mean of the delays of equivalent measurements, i.e. measurements performed in the same position on the sample. {\bfseries m-r}, Occurrence of measured delay values with histogram bins of \qty{5}{\as} width. The black curve indicates a normal distribution corresponding to the weighted mean value $\langle \tau \rangle$ (Eq. \ref{eq:Eq6} in Methods) and weighted standard deviation $\sigma$ (Eq. \ref{eq:Eq7} in Methods).}
		\label{fig:FigS2}
	\end{figure}
	
	\begin{figure}
		\centering
		\includegraphics[width = \textwidth]{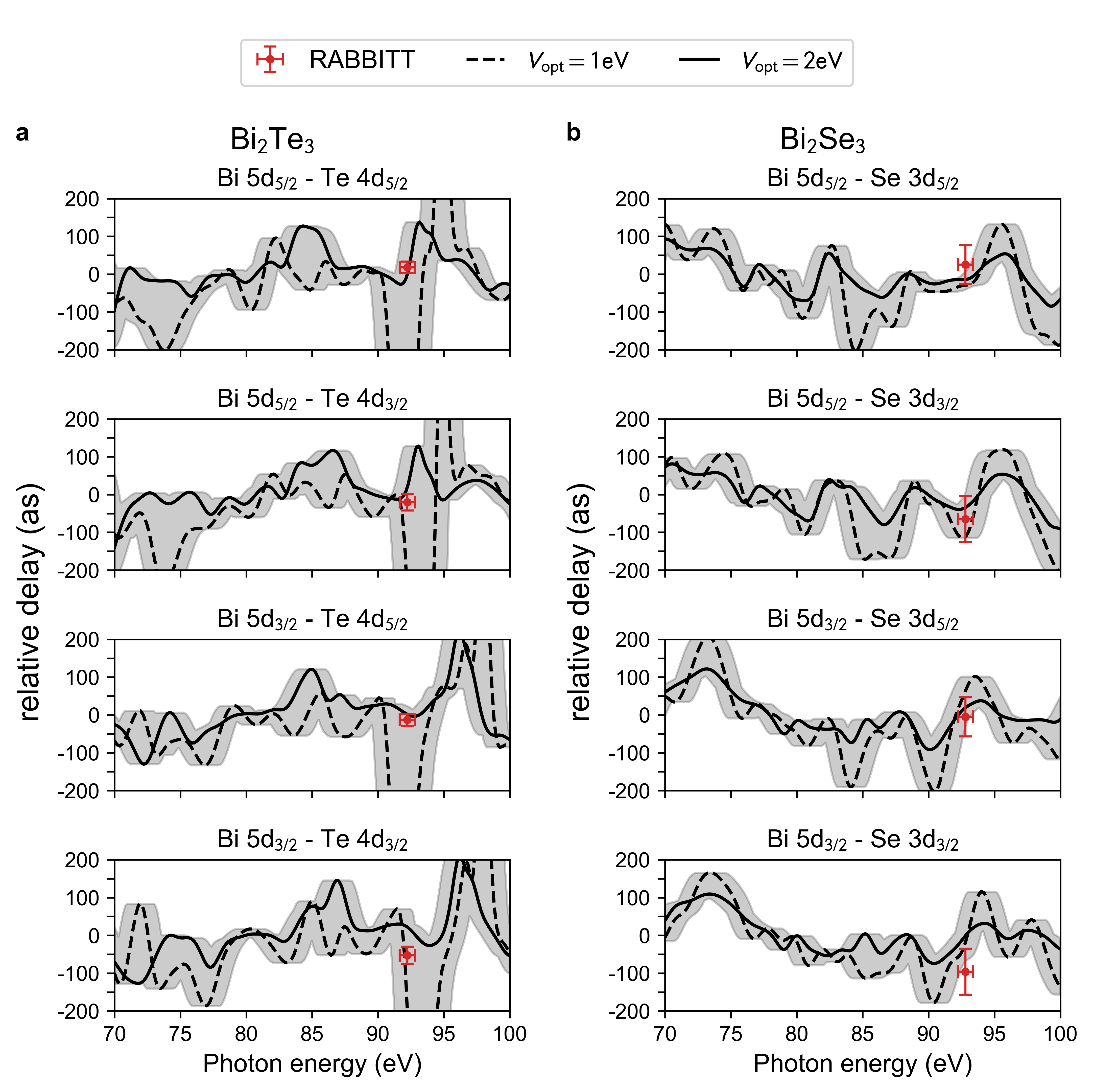}
		\caption{{\bfseries Comparison of measured non-differential relative photoemission delays with predictions based on OSTEWS theory.} {\bfseries a} and {\bfseries b}, Delays for Bi$_2$Te$_3$ and Bi$_2$Se$_3$, respectively. Data points correspond to the weighted mean value of all experimental RABBITT delays and the respective error bears are determined by the weighted standard deviation. Relative OSTEWS-delays are plotted as dashed ($V_{\mathrm{opt}} = \qty{1}{\eV}$) and solid lines ($V_{\mathrm{opt}} = \qty{2}{\eV}$) and the gray shaded area indicates a possible energy shift of $\pm \qty{1}{\eV}$ for each of the relative OSTEWS-delays.}
		\label{fig:FigS3}
	\end{figure}
	
	\begin{figure}
		\centering
		\includegraphics[width = 0.5\textwidth]{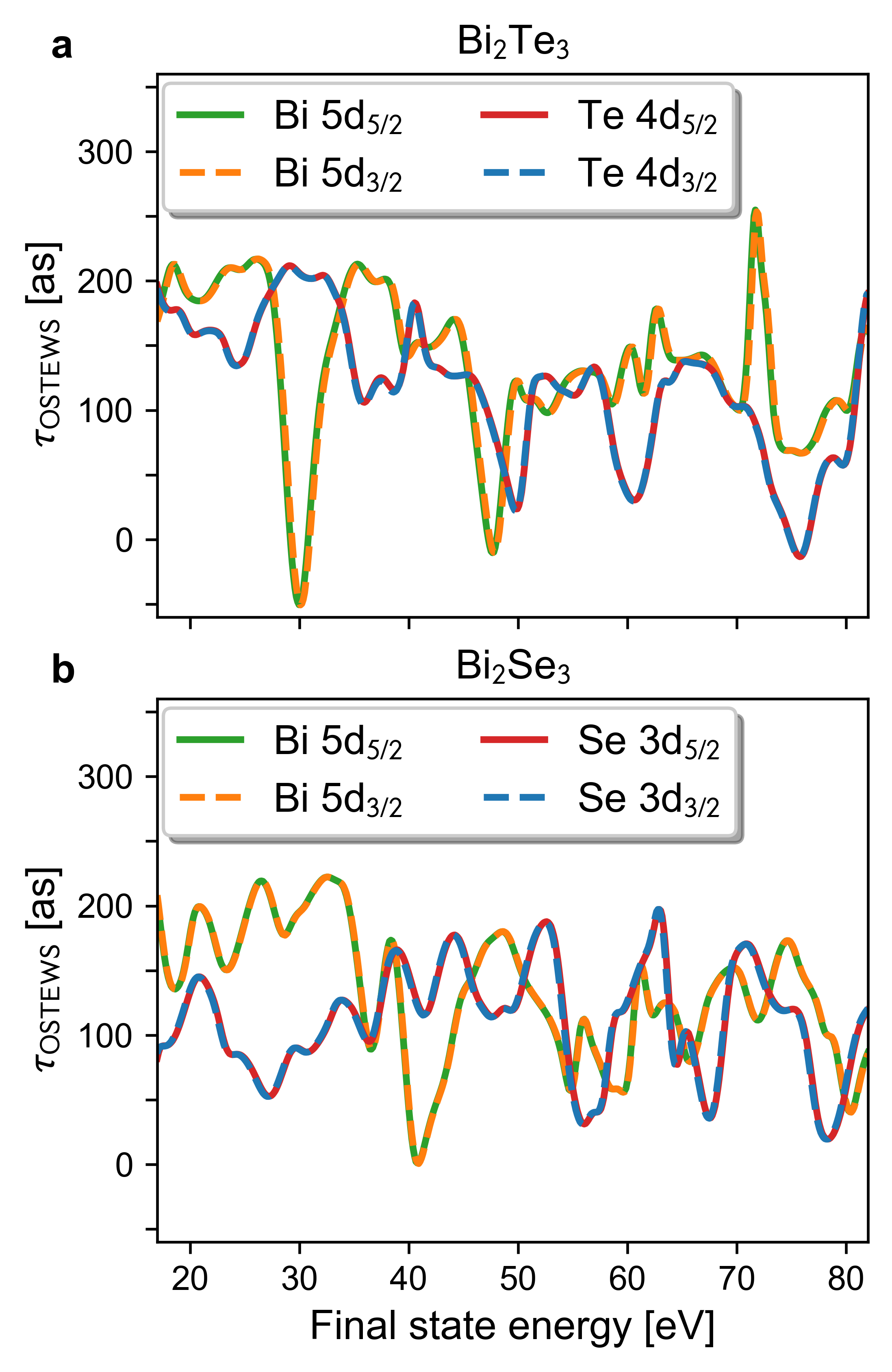}
		\caption{{\bfseries Absolute photoelectron escape time derived by one-step photoemission theory.} {\bfseries a} and {\bfseries b}, OSTEWS delays $\tau_{\mathrm{OSTEWS}}$ as a function of final state energy with respect to the valence band maximum for Bi$_2$Te$_3$ in part {\bfseries a} and Bi$_2$Se$_3$ in part {\bfseries b} for $V_{\mathrm{opt}} = \qty{2}{\eV}$. Note that the spin-orbit interaction of the initial states does affect the OSTEWS-delays only marginally since dashed and solid line almost perfectly coincide for Bi, Te and Se core level emission.}
		\label{fig:FigS4}
	\end{figure}

\end{document}